# A Verifiable Partial Key Escrow, Based on McCurley Encryption Scheme

Kooshiar Azimian[1,2]
azimian@ce.sharif.edu

Javad Mohajeri[1]
mohajeri@sharif.edu

Mahmoud Salmasizade[1]
salmasi@sharif.edu

Siamak Fayyaz[1,3]
fayyaz@ee.sharif.edu

[1] Electronic Research Center, Sharif University of Technology

[2] Department of Computer Engineering, Sharif University of Technology

[3] Department of Electrical Engineering, Sharif University of Technology

**Abstract.** In this paper, firstly we propose two new concepts concerning the notion of key escrow encryption schemes: provable partiality and independency. Roughly speaking we say that a scheme has provable partiality if existing polynomial time algorithm for recovering the secret knowing escrowed information implies a polynomial time algorithm that can solve a well-known intractable problem. In addition, we say that a scheme is independent if the secret key and the escrowed information are independent. Finally, we propose a new verifiable partial key escrow, which has both of above criteria. The new scheme use McCurley encryption scheme as underlying scheme.

# 1 Introduction

A key escrowed encryption scheme is a scheme in which a trusted third party (TTP) has the ability of decrypting cipher-texts in special circumstances. The goal is to retrieve important plain-texts in case of key loss or denial of a malicious user to decrypt a cipher-text

## 1.1 Partial key escrow (PKE)

The idea of partial key escrowing, which firstly presented by Shamir in 1995, is to deprive the TTP from decrypting the cipher-text immediately. Shamir proposed to let the TTP know the first 8 bits of the 56-bit key. Consequently, the TTP still needs to brute-force a key space of size $2^{48}$ to obtain the private key [Sh95]. It is easy to see that performing $2^{48}$ steps is not infeasible. However, it becomes hard to immediately and simultaneously uncover many keys.

**1.2 Verifiable partial key escrow (VPKE)**

When the TTP get escrowed information, he must be sure that the user does not cheat. That is, it should be possible to check that the secret can be found in the expected time using escrowed information. The issue of verifiability was introduced independently by Micali in [M95] and Bellare and Goldwasser in [BG95]. They proposed the first VPKE schemes with using Diffie-Hellman and RSA as underlying cryptosystem.

**2. The new issues**

In this section, we introduce the two new issues and discuss why they are important in concept of key escrow encryption schemes.

**2.1 Provable partiality**

Partiality is one of the main criteria of a proper key escrow scheme. As discussed in [Sh95], [M95] and [BG95] it causes the TTP cannot uncover many keys simultaneously.
In [M95] Micali proposed a key escrow scheme based on Diffie-Hellman and claimed that it is partial because the best-known algorithm to uncover the key using escrowed information needs more than $2^{40}$ steps. However, as presented in [WO96] there exist an algorithm can uncover the key in much less than $2^{40}$ steps. Generally, the most of VPKE give escrowed information to the TTP and the TTP attacks on the underlying scheme using that information. It supposed that the scheme has partiality if the proposed attack takes a relatively long time. However, it is possible that, there exist another attack on the scheme, which takes short time, with using the escrowed information.
 It is easy to see that, existing such an attack break the partiality of the scheme. To avoid such problem we introduce the new issue, provable partiality.
We say that a scheme is provably partial, if existing algorithm, which can break the scheme using escrowed information in poly-time, yields an algorithm, which can solve a well-known intractable problem in poly-time.

**2.2 Independency**

We say that a VPKE is independent if the escrowed information and the secret are independent. That is each user can change his secret without changing escrowed information. This property, causes each entity can change its secret every time without need to extra communication between the TTP and the entity. The new property makes the system more reliable and compatible.
Notice that changing the secret frequently causes the TTP cannot uncover the secret before receiving court order easily.

**3 The new scheme**

**3.1 McCurley encryption scheme**

The McCurley encryption scheme is an ElGamal like cryptosystem but it works in a subgroup of $Z_N^*$ where $N$ is a special-form composite number [Mc88]. In the proposed scheme each user $A$, produce a module $N = pq$, where $p$ and $q$ are two large primes having below criteria:

- $p = 3 \pmod 8$ and $q = 7 \pmod 8$
- $(p-1)/2$ and $(q-1)/2$ are primes
- $(p+1)/4$ and $(q+1)/8$ have large prime factors.

Then $A$ chooses a random number $S$, computes $y = 16^S \pmod N$. Finally $A$ makes $y$ and $N$ public and keeps $S$ secret.

For sending a message $m$ from each user $B$ to $A$, $A$ and $B$ must do the following:

- $B$ chooses a random number $k$, computes $u = 16^k \pmod N$, $t = m.y^k \pmod N$ and sends $(u,t)$ to $A$.
- $A$ can decrypt the cipher by computing $m = t(u^{-S}) \pmod N$

The proposed scheme is provably secure, based on intractability of factoring [Sh85, Mc88]

### 3.2 The new VPKE

The McCurley encryption scheme naturally has a nice property such that the user can keep secret, two independent information. One is the factoring of the module and another is $S$. In the new key escrow system we use this property. A prime factor of $N$ is given to the TTP, as escrowed information and $S$ will be the secret.

The TTP, knowing prime factors of $N$, can break the system by computing two discrete logarithms modulo $p$ and $q$. If the key size is chosen properly, the running time needed for computing the two discrete logarithms will be reasonable for uncovering the secret in a key escrow system [G93, Sh95]. Now, a 768-bit module seems to be proper one.

## 4 The new system advantages

The new system is a VPKE scheme because the TTP can verify the honesty of the user, by testing the equation $N = pq$. Moreover, the new scheme is a partial key escrow because the secret S, is not given to the TTP.

In this section, we show that the new system has both provable partiality and independency.

### 4.1 Provable partiality

From [Mc88] we know that existing of a poly-time algorithm, which can break the system even knowing the factorization of $N$, will yields a poly-time algorithm, which can solve the Diffie-Hellman problem modulo a prime. Thus, the new system is a provably partial system, based on intractability of Diffie-Hellman.

### 4.2 Independency

It is easy to see that the escrowed information –factorization of the module- and the secret $S$ are independent in the new system. Thus without changing $N$, the user can change $S$ any time he wants. In former systems such [Sh95], [M95] and [BG95], the escrowed information and the secret related to each other and the user may not change the secret without communicating with the TTP.

## 5 Conclusion

In this paper, we proposed two new concepts concerning the notion of key escrowing. We showed these two new properties make a key escrow system more reliable and compatible. Then we introduced a new VPKE system, which satisfies the both properties. The new system use McCurley encryption scheme as underlying scheme. As discussed in section 4, the new scheme has high security and reasonable secret-recovery time.